\documentclass[structabstract]{aa}  
\usepackage{graphicx}
\usepackage{txfonts}
\usepackage{natbib}
\bibpunct{(}{)}{;}{a}{}{,}
\begin{document}
   \title{Narrow Band H$\alpha$ Photometry of the Super-Earth GJ 1214b with GTC/OSIRIS Tunable Filters}

   \author{F. Murgas \inst{1}
          \and
          E. Pall\'{e} \inst{1}\fnmsep \inst{2}
          \and
	       A. Cabrera-Lavers \inst{1}\fnmsep \inst{2}
	      \and
	       K. D. Col\'{o}n \inst{3}
	      \and
	       E. L. Mart\'{i}n \inst{4}
	      \and
	      H. Parviainen \inst{1}
          }

   \institute{Instituto de Astrof\'{i}sica de Canarias (IAC), E-38205 La Laguna, Tenerife, Spain\\
              \email{murgas@iac.es}
         \and
             Departamento de Astrof\'{i}sica, Universidad de La Laguna (ULL), E-38206 La Laguna, Tenerife, Spain
         \and
             Department of Astronomy, University of Florida, Gainesville, FL 32611, USA
         \and
          	 Centro de Astrobiolog\'{i}a (CSIC-INTA), Ctra. Ajalvir km. 4, 28850 Torrej\'{o}n de Ardoz, Madrid, Spain
             }

   \date{Received XXXX XXXX, 2012; accepted XXXX XXXX, 2012}

% \abstract{}{}{}{}{} 
% 5 {} token are mandatory

  \abstract
   {}
   {The super-earth planet GJ 1214b has recently been the focus of several studies, which use the transit spectroscopy technique to determine the nature of its atmosphere. Here, we focus on the H$\alpha$ line as a tool to further restrict the nature of GJ 1214b's atmosphere.}
   {We used the Gran Telescopio Canarias (GTC) OSIRIS instrument to acquire narrow-band photometry with tunable filters. We were able to observe the primary transit of the super-Earth GJ 1214 in three bandpasses: two centered in the continuum around H$\alpha$ (653.5 nm and 662.0 nm) and one centered at the line core (656.3 nm). We measure the depth of the planetary transit at each wavelength interval.}
   {By fitting analytic models to the measured light curves, we were able to compute the depth of the transit at the three bandpasses. Taking the difference in the computed planet-to-star radius ratio between the line and the comparison continuum filters, we find $\Delta (R_p/R_\star)_{H\alpha-653.5} = (6.60 \pm 3.54) \times 10^{-3}$ and $\Delta (R_p/R_\star)_{H\alpha-662.0} = (3.30 \pm 3.61) \times 10^{-3}$. Although the planet radius is found to be larger in the H$\alpha$ line than in the surrounding continuum, the quality of our observations and the sigma level of the differences (1.8 and 1.0, respectively) do not allow us to claim an H$\alpha$ excess in GJ 1214's atmosphere. Further observations will be needed to resolve this issue.}
   {}


   \maketitle
%
%________________________________________________________________

\section{Introduction}

Exoplanetary research is one of the fastest-growing areas of astronomy. Particularly exciting was the detection of transiting planets with a mass lower than Neptune (e.g., \citealp{Leger2009}, \citealp{Howard2011}), which has opened up the study of planets with masses in the range 1.5 to 10 M$_{\oplus}$, the so-called super-Earths. Even more recently, Kepler results have pushed the detection limit down to Mars-size planets (KOI 961.01, 961.02, and 961.03; \citealp{Muirhead_2012}). Super-Earths are important because they have a mass range that is not present in the solar system, they can bring new clues about planetary formation, and they are good candidates for future searches for life. 
   
   One of the most-studied super-Earths is GJ 1214b, which was discovered by \citet{Charbonneau2009} as part of a ground-based exoplanet search in M dwarf stars (MEarth, \citealp{NutzCharb_2008}). Since this system has a larger planet-to-star radius ratio ($R_p/R_\star$) than most of the super-Earths detected so far (e.g., Corot 7b, Kepler 22b), GJ 1214b has become a very interesting candidate for the study of its atmospheric composition. According to some simulations (\citealp{RogersSeager_2010}), there is a degeneracy in the models that account for the possible atmospheric composition of GJ 1214b based on its density and irradiation. Thus, the planet can either be composed of a) a rocky/ice core surrounded by a primordial atmosphere dominated by hydrogen and with a relatively large-scale height, b) a water and ice core (Waterworld) with an atmosphere dominated by water vapor with a small-scale height, or c) GJ 1214b can be a rocky planet with an atmosphere formed by outgassing with a large atmospheric scale height. 
   
   Transmission spectroscopy has proven to be a successful technique to constrain the planetary composition and even detect molecules in extra-solar planets (e.g., \citealp{Charbonneau2002}; \citealp{Tinetti2007}, \citealp{Sing2011}). The methodology is based on the fact that each element and molecule present in the planetary atmosphere possesses a different opacity, thus producing a slight change in the observed radius of the planet $R_p$ (therefore changing the ratio $R_p/R_\star$) at different wavelengths. By comparing the measured changes in $R_p/R_\star$ with the predictions made by different atmospheric models, one can infer the composition of the planet.
   
   In the case of GJ 1214b, several studies have been made to try to constrain its atmospheric composition through transmission spectroscopy (e.g., \citealp{Bean2010}, \citealp{Desert2011b}, \citealp{Crossfield2011}, \citealp{Croll2011}), and disentangle the degeneracy between the models that account for its density. Using spectra taken with VLT/FORS, \cite{Bean2010} did not find evidence of any significant features between 780 nm and 1000 nm. This result has been confirmed by \cite{Bean2011} and other studies (e.g., \citealp{Berta2012}; \citealp{deMooij2012}) that did not find evidence of features present in the atmosphere of GJ 1214b. This lack of features is interpreted as evidence for the presence of a metal-rich atmosphere and/or thick clouds that produce a constant absorption across a broad range in wavelength. 
  
  Here, we present a new set of narrow-band H$\alpha$ photometry of a transit of GJ 1214b on August 17 2011, taken with the Optical System for Imaging and low Resolution Integrated Spectroscopy (OSIRIS, \citealp{Cepa2000}) at the 10.4 m Gran Telescopio Canarias (GTC), located at Observatorio Roque de los Muchachos in La Palma, Canary Islands, Spain. In this study of the atmospheric signature of a super-Earth, we use a different approach than in the literature; instead of using broadband photometry or low-resolution spectroscopy, we use narrow-band imaging with tunable filters.

%%%%%%%%%%%%%%%%%%%%%%%%%%%%%%%%%%%%%%%%

\section{Data}
\subsection{Observations}

The OSIRIS instrument consists of a mosaic of two Marconi CCD detectors, each with $2048 \times 4096$ pixels and a total unvignetted field of view of $7.8 \times 7.8$ arcmin, giving a plate scale of 0.127 arcsec/pix. For our observations, we choose the $2 \times 2$ binning mode with a readout speed of 200 kHz (that has a gain of 0.95 e-/ADU and a readout noise of 4.5 e-), defining a single window of $1.45 \times 6.87$ arcmin to increase the sampling in the sequence. This configuration produced a readout overhead time of only 2.9 s. 

We used the OSIRIS Red Tunable Filter (RTF) for our observations, which consists of a Fabry-Perot etalon that allows for narrow-band imaging, selecting both the central wavelength and the width of the filter. In the case of OSIRIS, the user can select a filter centered in the wavelength range of 651-934.5 nm, with a width of 1.2-2.0 nm. The use of tunable filters (TFs) in exoplanetary science has been explored in recent years (e.g., \citealp{Colon2010}, \citealp{Sing2011}, \citealp{Colon2011}). The collecting capability of the 10.4 m GTC telescope in combination with the flexibility of the TFs makes it possible to obtain transit depths at a fixed wavelength with enough accuracy to detect possible atomic signatures in planet atmospheres. This was the case with the first detections of potassium in the extrasolar planet XO 2b (\citealp{Sing2011}) and the study of the same element in HD 80606b (\citealp{Colon2012}).

\begin{figure}
   \centering
  \includegraphics[width=\linewidth]{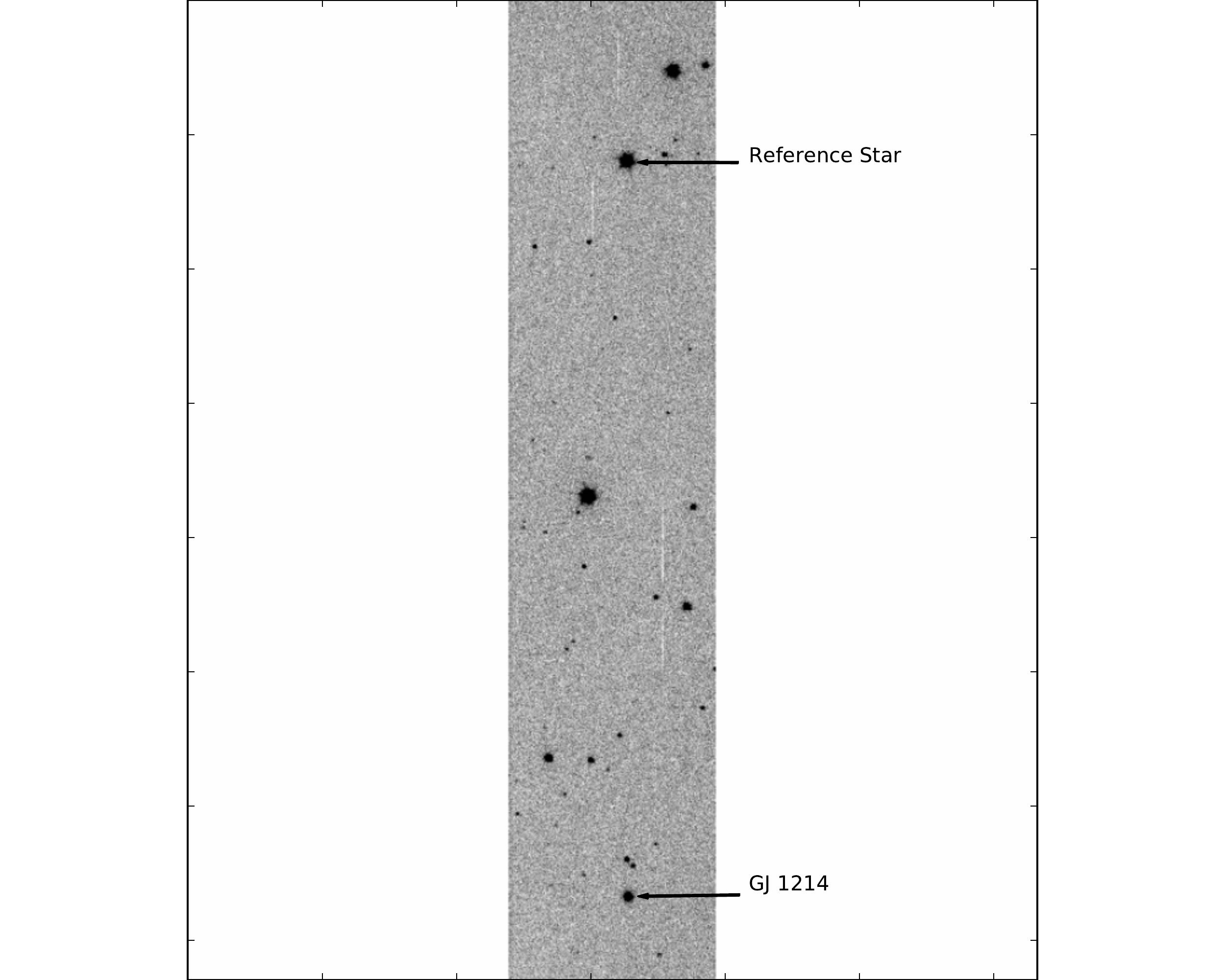}
      \caption{Raw H$\alpha$ GTC/OSIRIS image. The position of GJ 1214 and the reference star used in the photometry is shown.}
         \label{Fig1}
    \end{figure}

As in any Fabry-Perot etalon, each pixel in the CCD in OSIRIS RTF images has a slightly different wavelength, increasing radially from the optical center. This increase thus produces ``rings'' of constant wavelength with respect to the RTF center located nearly in the middle of the detector system. As a result, stars with different positions will have different observed wavelengths, unless they are at the same distance from the optical center. For this reason, the field of view of OSIRIS was rotated 149.63 deg (North is down-right in Figure \ref{Fig1}) in order to have GJ 1214 and one bright reference star located at the same distance, about 3.05 arcmin, from the optical center. In this way, they can be at the same wavelength, while the rest of the stars are at different distances from the optical center and hence at different wavelengths. At the distance of 3.05 arcmin from the optical center the tunable filter was set up at 658.0 nm, 661.2 nm, and 666.9 nm to get the desired wavelengths of 653.5 nm, 656.3 nm, and 662.0 nm at both the target and reference star, respectively. For each narrow filter, we used a width of 1.2 nm and we set an exposure time of nine seconds for each band. This gave a duty cycle of more than 75\%, taking into account the small readout time used along the series. The continuum-line tunings were cycled during the observation using increasing wavelength (i.e., 658.0 nm - 661.2 nm - 666.9 nm). Figure \ref{Fig1} presents one of the raw OSIRIS windowed images of GJ 1214 obtained in the series, showing the position of the target and reference star used in the photometry. 
  
Data for GJ 1214 were taken in service mode on August 17, 2011. Transit observations began at 22:50 UT and ended at 1:00 UT, during which the airmass ranged from 1.27 to 2.37. The observing conditions were good, with a clear sky and bright moon. Seeing was good and stable (0.7-0.8 arcsec) along the whole transit, so that a slight defocus was needed to avoid saturation in the reference star. This defocus was produced by changing the nominal position for GTC secondary mirror by $\pm$0.1 mm and maintained without changes for the three different wavelengths.  With this procedure, a maximum count level of around 35,000 - 40,000 ADUs was ensured along the observations at the reference star, the brightest star in the field, well below the saturation level (65,000 ADUs) for OSIRIS CCDs. At 23:45 UT, near mid transit, some guiding problems (caused by telescope defocus) resulted in the loss of some images and a slight change in the position of the stars (about eight pixels) for the rest of the observations. This change, due to the decrease of the radial wavelength outwards from the optical center, translates to a difference of less than 0.05 nm with respect to the initial wavelength setup and thus has no effect in the photometric measurements. The RTF also suffers from a calibration effect that depends on the rotator angle (see OSIRIS user manual \footnote{www.gtc.iac.es/en/pages/instrumentation/osiris.php}). However, for the range of rotator angles covered along the sequence, the differences in the TF tuning for this effect were again lower than 0.05 nm, so that no recalibrations were needed during the transit.
   
\subsection{Data reduction and Photometry}

Photometric reduction was completed using standard procedures. Bias and flat-field images for each filter were processed using the Image Reduction and Analysis Facility (IRAF\footnote{IRAF is distributed by the National Optical Astronomy Observatories, which are operated by the Association of Universities for Research in Astronomy, Inc., under cooperative agreement with the National Science Foundation.}) routines, while the bias, flat-fielding correction, centroid computation, and aperture photometry were performed using specific Interactive Data Language (IDL) scripts. 

Finally, only one star was used as reference for the differential photometry, since we could only get one bright star at exactly the same distance as GJ 1214 from the optical center (that is, at the same wavelength as GJ 1214). Although tests were made to compute the flux ratio between GJ 1214 and other reference stars that were available within the OSIRIS field of view (FOV), we concluded that our initial reference star was the best option because it delivered better results in rms than an ensemble of stars. 

Due to the wavelength variation across the OSIRIS RTF FOV, ring-like structures resulting from sky emission lines appear in the images. To remove any noise that these structures might introduce to the photometry, we used the IRAF package which was created to reduce images taken with tunable filters at the Anglo-Australian Telescope TFRed\footnote{http://www.aao.gov.au/local/www/jbh/ttf/adv\_ reduc.html} (\citealp{Jones2002}). This artificially dithers each frame in order to create a sky image, which is then substracted from the original data. The use of TFRed, however, did not improve the quality of the photometry, as noted in previous works (\cite{Colon2010}), so we decided to follow the traditional approach of performing aperture photometry by using a fixed aperture and a sky substraction based on the average counts in an annulus away from the star. The aperture and sky ring combination that delivered the lowest rms was an aperture of 2.25 arcsec (approximately three times the measured FWHM of the objects) and a sky level computed using a ring of 2.5 arcsec of width that started five arcsec away from the center of the stars. The errors in the photometry were computed by the IDL routine APER\footnote{http://idlastro.gsfc.nasa.gov/} which uses photon count statistics to estimate the error level. However, the computed error was smaller than the point-to-point spread observed in the light curve, so we adopted the rms for the out-transit data as final error estimate.

After computing the flux ratio between the target and the reference star, a second-degree polynomial was fitted to the out-of-transit data points in order to correct a trend in the light curve. This trend is probably produced by atmospheric extinction affecting the flux of the stars in a different way due to their distinct color. Using the correlation between the flux ratio and the airmass, another polynomial fit was made as a test. However, since it delivered similar results in terms of rms of out-transit data and curve shape as using the temporal trend did,  we decided to use the initial correction for the light curve presented here.

We obtained the universal time of data acquisition for each frame by using the recorded headers of the images, which indicate the opening and closing time of the shutter, in order to compute the time of mid exposure. Then using the code written by \cite{Eastman2010}\footnote{http://astroutils.astronomy.ohio-state.edu/time/}, we computed the barycentric julian date in barycentric dynamical time (BJD) using the mid-exposure time for each of the science images to produce the final light curves at the three different wavelengths analyzed here. 

%%%%%%%%%%%%%%%%%%%%%%%%%%%%%%%%%%%%%%%%

\section{Data Analysis}
\subsection{Light Curve Fitting}

The light curves obtained for each of the three filters, together with the best-fitting synthetic models, are shown in Figure \ref{Fig2}. As pointed out, the telescope temporarily lost the guiding near mid-transit, producing a change in the position of the stars and a few outliers in the photometry, which are more noticeable in the 656.3 nm and 662.0 nm filters.

To fit the observed light curve to synthetic models, we fixed two of the orbital elements, the ratio $a/R_\star = 14.9749$ and the inclination of the system $i=88.94^\circ$ (both taken from \citealp{Bean2011}) and also assumed an eccentricity of $e=0$. Since GJ 1214b did not show any evidence of central transit time variations (\citealp{Bean2010}, \citealp{Berta2012}), we fixed the period to $P=1.58040481$ days (\citealp{Bean2011}). The central time of the transit ($T_0$) was left as a free parameter for each of the three filters and so that we could see if there was considerable departure from the predicted $T_0$. The predicted central time of $T_{Eph}=2455791.49643382$ BJD was calculated using a linear ephemeris (with $T_0$ taken from \citealp{Bean2011}). Since our data consists of observations taken with very narrow filters, we considered that fixing the limb-darkening (LD) coefficient value extrapolated from the results found for broadband photometry could introduce a source of error in the estimation of the transit depth. Thus, we decided to also leave the LD coefficient as a free parameter in the fitting process. 

The models of \citet{Gimenez06} were fitted separately for each filter, leaving as free parameters the central time of the transit, LD coefficient, and the depth of the transit. Here we have chosen to use a linear LD law, instead of a quadratic or nonlinear coefficients, because the data is noisier than a priori expected based on telescope aperture. Thus, we consider that the use of a linear LD law is enough to account for the characteristics of the curve.

\subsection{Parameter Estimation}
In this section, we describe the bayesian method used to estimate the fitted parameters to the light curve. The procedure is similar to the one used, and amply detailed, in \cite{Berta2012}.

The first step to estimate the distribution of the free parameters chosen to fit the observed light curve (depth and LD) is to minimize the $\chi^2$ function that compares the data with the synthetic models. It is:

\begin{equation}
\chi^2 = \sum_{i=1}^N \frac{(d_i - m_i)^2}{\sigma^2} \, ,
\label{eq:chi2}
\end{equation}

where $d_i$ is the observed point of the light curve, $m_i$ the point taken from the comparison model, and $\sigma$ the error in the observation. Once  $\chi^2$ is minimized, we have an initial guess for the transit depth and LD coefficient to feed to our bayesian analysis. 
 
In the bayesian analysis, the Metropolis-Hasting method is applied to find the distribution of probabilities of the transit depth and LD parameters. The algorithm compares the observations ($D$) with different synthetic models ($M$), which were computed using the transit depth and LD coefficient generated randomly with a probability distribution. In our case, and for the parameters mentioned before, we adopted a normal random distribution centered in the value found by the $\chi^2$ minimization as well as a standard deviation found by trial and error to see which combination of numbers decreased the time of convergence of the algorithm. In each step, a comparison is made using Bayes' theorem to see if the model in the $t+1$-th iteration ($M_{t+1}$) has higher probability than the previous step ($M_t$). This is done by evaluating their ratio of probabilities $r$ given by

\begin{equation}
r = \frac{ P(D\vert M_{t+1}) }{ P(D\vert M_{t})}  \frac{P(M_{t+1})}{P(M_t)} \, ,
\label{eq:bayes}
\end{equation}

where $P(D\vert M)$ is the posterior likelihood of the model, which evaluates how good our choice of parameters is compared to the data. In our case, we computed the likelihood as in \cite{Berta2012}:

\begin{equation}
\ln P(D\vert M) = -N \ln(s\sigma) - \frac{\chi^2}{2s^2} + constant \, ,
\label{eq:likeli}
\end{equation}

where $N$ is the number of data points, $\chi^2$ is described by Equation \ref{eq:chi2}, and $s$ is a parameter introduced to take into account the true noise level in our data. In each iteration, we compute Eq. \ref{eq:likeli} using a normal random distribution to generate a new value for $T_0$, $s$, the transit depth, and the LD coefficient.

Called the prior probability distribution, $P(M)$ describes the knowledge that we have of the space in which our parameters are distributed. For the central time ($T_0$), depth ($k$), and $s$ parameter, we use Jeffreys prior and for the LD coefficient ($c$) uniform prior:

\begin{eqnarray}
 P(M) & = &  \frac{1}{T_0\ln(T_{0max}/T_{0min})} \frac{1}{k\ln(k_{max}/k_{min})} \frac{1}{ s\ln(s_{max}/s_{min}) }  \nonumber\\
     & \times &  \frac{1}{(c_{max} - c_{min})}  \, , \nonumber\\
\label{eq:prior}
\end{eqnarray}

where $x_{min}$, $x_{max}$ with $x=[T_0,k,c,s]$ represents the limits of the domain of the central time, depth, LD, and $s$ respectively. For the central time, we use [$T_{Eph}$-0.001042,$T_{Eph}$+0.001042], for $k$ [0.008,0.5], for LD [0.1,1.0], and for $s$ [0.5,2.0].

Once we have computed the ratio $r$, we compare it to a number taken from a random uniform distribution $U$ in the interval [0, 1]. If $r \geq U(0,1)$, we accept the present model and set $M_t=M_{t+1}$. To ensure that the method converges to the distribution of probabilities for each fitted parameter, this process is repeated several times. In our case, we performed $2.5\times 10^5$ iterations for each light curve.

Since the algorithm takes time to start sampling the true distribution of the parameters, we have to eliminate some of the first iterations in a process known as a \textit{burn-in} period.  For the depth, central time, LD coefficient, and $s$, we did not consider the first $2\times 10^4$ iterations. By computing the autocorrelation for each fitted parameter, we were able to obtain the correlation length of the chain and keep only the parameter values that were independent in order to establish their probability distribution. At the end, for each parameter we have of the order of $10^4$ points to compute their distribution.
 
\begin{figure}
   \centering
   \includegraphics[width=\linewidth]{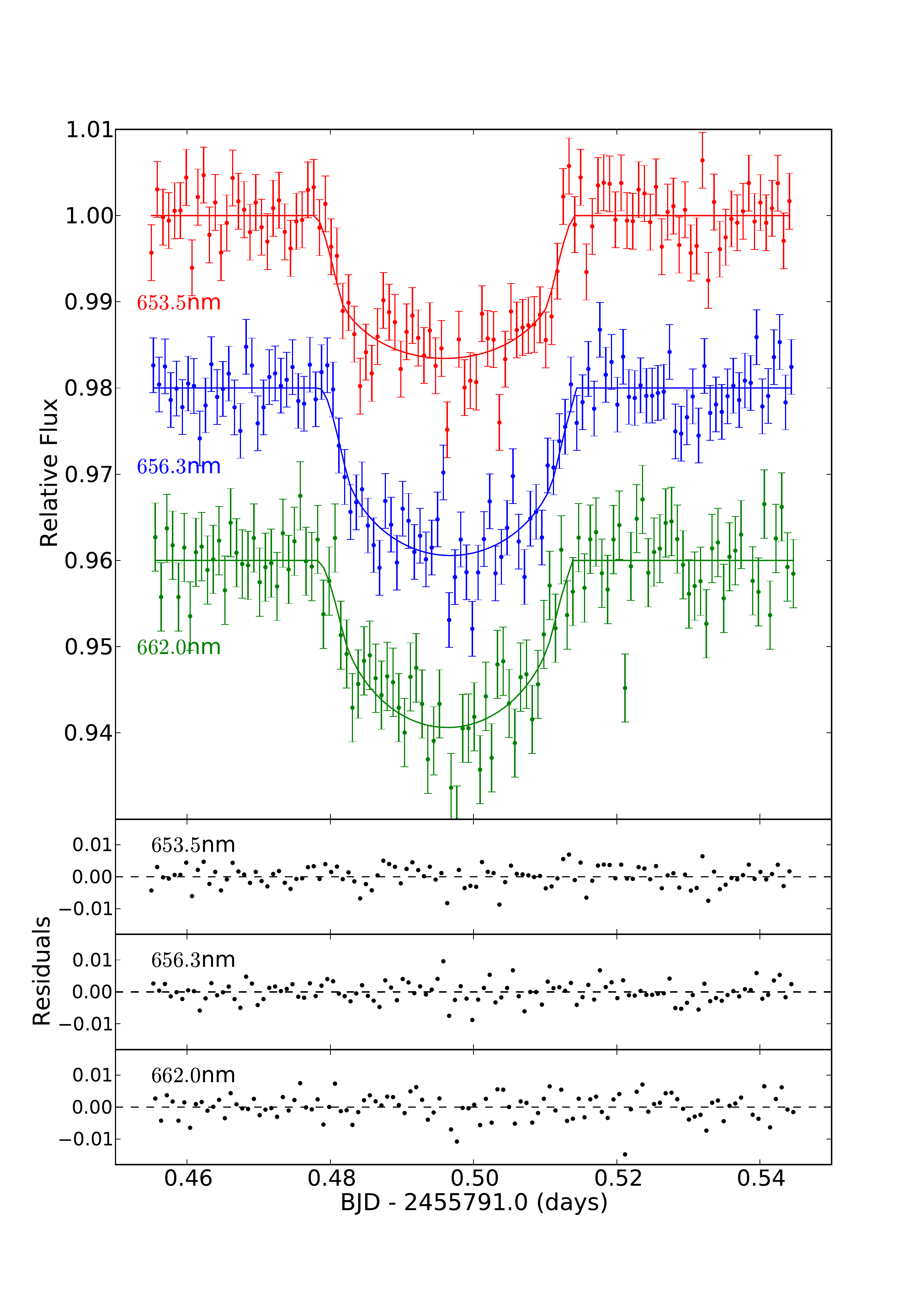}
    \caption{GJ 1214b transit light curves obtained with GTC/OSIRIS tunable filters. In the bottom panels, the residuals of the curve fit are shown for each filter.}
     \label{Fig2}
    \end{figure}

Once we obtained the probability distribution of the parameters, we computed the percentiles of the distribution in order to obtain the median and standard deviation for each parameter. The median is the value below which 50\% of the observed points are found, while the standard deviation was obtained using the value by which 68.27\% of the data is found (around the median). By using this procedure, we are able to find the median and standard deviation for a nonsymmetric function, although most of the fitted parameter distibutions present gaussian shapes. In Table \ref{tabl1}, we give the results of the light curve fitting and parameter estimations for the three different filters. The values and quoted errors are the median and standard deviation derived from the probability distribution, respectively. We used the median values for the central time, transit depth, and LD coefficient to produce the model curves and the $s$ parameter to estimate the true level of error by $e_{true}=s\sigma$ (see Figure \ref{Fig2}). Looking at Table \ref{tabl1}, we note that the LD coefficients computed have consistent values for all the filters within the relative large errors, which they seem to increase to shorter wavelengths.

\begin{table}
\caption{GJ 1214b central transit times, depths, and linear limb-darkening coefficients from GTC/OSIRIS data.}
\label{tabl1}
\centering
\begin{tabular}{c c c c} 
\hline\hline                 
Filter & $R_p/R_\star$  & $C_1$ & $\Delta T_0$ (days)\tablefootmark{a} \\ 
\hline                       
 653.5 nm &  0.1151 $\pm$ 0.0025  &  0.670  $\pm$ 0.143 &   0.00040 $\pm$ 0.00031 \\
 656.3 nm &  0.1217 $\pm$ 0.0025  &  0.793  $\pm$ 0.122 & -0.00036 $\pm$ 0.00035 \\
 662.0 nm &  0.1184 $\pm$ 0.0026  &  0.917  $\pm$ 0.099 &   0.00007 $\pm$ 0.00044\\
\hline
\end{tabular}
\tablefoottext{a}{Offset $\Delta T_0=T_{Eph}-T_0$ computed from ephemeris central time $T_{Eph}=2455791.49643382$ BJD.}
\end{table}

The rms of the residuals from the best fit for each light curve (Fig. \ref{Fig2}) are 3900 ppm (653.5 nm), 3100 ppm (656.3 nm) and 3200 ppm (662.0 nm).

As an additional test, we performed a prayer bead analysis (\citealp{Moutou2004}, \citealp{Gillon2007}, \citealp{Desert2011}) in each filter (separately) to get another estimation of the fitted parameter median values and errors and to see the possible impact of systematics in the fitting process. The test consisted of shifting the residuals from the initial fit (in this case using the values from Table \ref{tabl1}), adding these residuals to the initial fit, and adjusting the resultant light curve again; this process is repeated $N$ times, where $N$ is the number of observed data points.

For all the adjusted parameters, the median values found by the prayer bead test were consistent with the ones found by the bayesian approach, meaning that these parameters were not strongly affected by systematic errors. The errors computed using this method were smaller than the ones previously found, so we adopted as final error values the ones found by the bayesian approach.

%%%%%%%%%%%%%%%%%%%%%%%%%%%%%%%%%%%%%%%%

\section{Discussion and Conclusions}

It is easily seen in Figure \ref{Fig2} that the light curves show a fair degree of variability throughout the observations. Given the good quality of the weather and sky conditions during the observations, this noise may have arisen from an imperfect correction of atmospheric effects due to the use of only one reference star located relatively far from the target ($\sim$ 5.5 arcmin) and the high airmass ($>2$) at the end of the observations. This photometric dispersion could also have been produced by stellar activity; GJ 1214 has previously shown evidence of some level of activity in the form of spots and low-energy flares (\citealp{Kundurthy2011}), although it was first classified as an inactive star (\citealp{Hawley_1996}). This phenomenon could have affected our photometry since we focused on the H$\alpha$ line, which has been used as an indicator of stellar activity. The same effect could be present in the reference star used in the differential photometry.

\begin{figure}
   \centering
   \includegraphics[width=\linewidth]{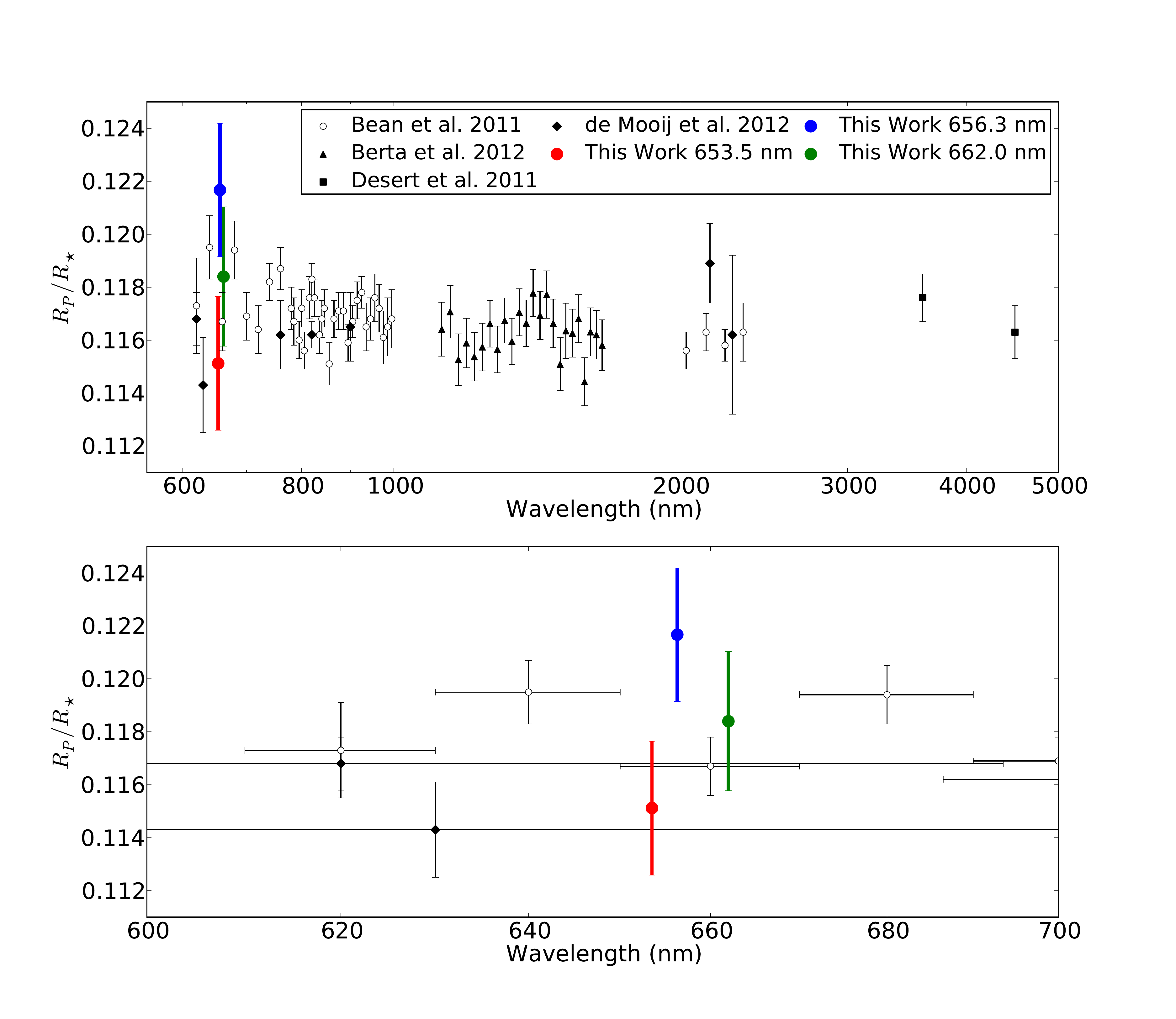}
      \caption{GJ 1214b transmission spectra obtained with GTC/OSIRIS tunable filters compared with other works. Top panel: full range of observed wavelengths by different authors, Bottom panel: optical observations (600 nm - 700 nm); the horizontal bars represent the width of the filters used in each observation.}
         \label{Fig3}
    \end{figure}

In Fig. \ref{Fig3}, we compare our results with previous works on GJ 1214b. It shows how well this planet has been studied since its discovery, and it possesses one of the most complete transmission spectra to date. Our data presents a bigger scatter when compared with the work of \cite{Bean2011}, which gave optical and near-infrared data taken at several facilities. The optical observations were made with the VLT FORS spectrograph, where the spectrum was binned into 20 nm wide wavelength intervals (a factor of 16.66 coarser spectral resolution than our RTF data), spanning 610 nm to 850 nm. The interval that contains our three data points (650 nm - 670 nm) presents a planet-to-star ratio of $R_p/R_\star = 0.1167 \pm 0.0011$, which differs from our value for the H$\alpha$ line of $R_p/R_\star =  0.1217 \pm 0.0025$. However, if we consider our two RTF setups in the continuum near the H$\alpha$ line, the retrieved planet-to-star radius ratio is fully consistent with the value found by \cite{Bean2011} within the error bars.

If we take the difference in the computed planet-to-star radius ratio between the line and the comparison continuum filters, we find that $\Delta (R_p/R_\star)_{H\alpha-653.5} = (6.60 \pm 3.54) \times 10^{-3}$ and $\Delta (R_p/R_\star)_{H\alpha-662.0} = (3.30 \pm 3.61) \times 10^{-3}$. The comparison between $H\alpha$ while the blue continuum (653.5 nm) is consistent at $1.8 \sigma$ and between $H\alpha$ and the red continuum (662.0 nm) is consistent at $0.9 \sigma$ level. As a result, we did not find a statistically significant detection of $H\alpha$. Still, the larger-than-expected planet radius in the H$\alpha$ line probably deserves new observations, aiming at higher SNR, to finally solve this issue.

This nondetection of $H\alpha$ is coherent with previous lower resolution results that do not find evidence of features presented in the atmosphere of GJ 1214b. In addition, our results can be used to constrain limits on the transmission spectrum GJ 1214b at high resolution.

%%%%%%%%%%%%%%%%%%%%%%%%%%%%%%%%%%%%%%%%

\begin{acknowledgements}

Based on observations made with the Gran Telescopio Canarias (GTC) in the Spanish Observatorio del Roque de los Muchachos of the Instituto de Astrof\'{i}sica de Canarias, on the island of La Palma. F. Murgas and H. Parviainen are supported by RoPACS, a Marie Curie Initial Training Network funded by the European Commission's Seventh Framework Program. E. Pall\'{e} acknowledges support from the Spanish MICIIN, grant \# CGL2009-10641. This material is based upon work supported by the National Science Foundation Graduate Research Fellowship under Grant No. DGE-0802270. E.L. Martin acknowledges support from the Consolider-Ingenio GTC project and from the PNAYA project AYA2010-21308.
\end{acknowledgements}

%%%%%%%%%%%%%%%%%%%%%%%%%%%%%%%%%%%%%%%%%%%%%%%%%%%%%%%%%%%%%%%%%%%%%%%

\bibliographystyle{aa}
\bibliography{biblio.bib}

%%%%%%%%%%%%%%%%%%%%%%%%%%%%%%%%%%%%%%%%%%%%%%%%%%%%%%%%%%%%%%%%%%%%%%

\end{document}